\documentclass[aps,prd,superscriptaddress,amsfont,graphicx,nofootinbib,preprintnumbers]{revtex4}%
\usepackage{color,graphicx,epsfig}
\usepackage{ifpdf}
\usepackage{amsmath}
\usepackage{bm}
\usepackage{color}
\usepackage[english]{babel}
\usepackage{graphicx}%
\usepackage{amsfonts}%
\usepackage{amssymb}
\usepackage{braket}
\usepackage{hyperref}

\definecolor{nicered}{rgb}{0.7,0.1,0.1}
\definecolor{nicegreen}{rgb}{0.1,0.5,0.1}
\hypersetup{colorlinks,citecolor= nicegreen,linkcolor= nicered}

\newenvironment{Eqnarray}{\arraycolsep 0.14em\begin{eqnarray}}{\end{eqnarray}}
\def\beqa{\begin{Eqnarray}}
\def\eeqa{\end{Eqnarray}}
\newcommand{\no}{\nonumber}

\newcommand{\beq}{\begin{equation}}
\newcommand{\eeq}{\end{equation}}
\newcommand{\bea}{\begin{eqnarray}}
\newcommand{\eea}{\end{eqnarray}}

\def\lsim{\mathrel{\rlap{\lower4pt\hbox{\hskip1pt$\sim$}}
     \raise1pt\hbox{$<$}}}         %less than or approx. symbol
\def\gsim{\mathrel{\rlap{\lower4pt\hbox{\hskip1pt$\sim$}}
     \raise1pt\hbox{$>$}}}         %greater than or approx. symbol

\begin{document}

\vskip1.5cm
%\title{Lessons from the LHCb measurement of CP violation in $B_s\to K^+K^-$}
\begin{center}
{\Large \bf \boldmath $b\to c\tau\bar\nu_{e,\mu}$ contributions to $R(D^{(*)})$}
\end{center}
\vskip0.2cm

\begin{center}
Shikma Bressler, Federico De Vito Halevy and Yosef Nir\\
%\today
\end{center}
\vskip 8pt

\begin{center}
{ \it Department of Particle Physics and Astrophysics,\\
Weizmann Institute of Science, Rehovot 7610001, Israel} \vspace*{0.3cm}

{\tt  shikma.bressler,federico.devitohalevy,yosef.nir@weizmann.ac.il}
\end{center}

\vglue 0.3truecm

%\begin{abstract}
  \vskip 3pt \noindent
\centerline{\large Abstract}
The $R(D^{(*)})$ puzzle stands for a $\sim3\sigma$ violation of lepton flavor universality between the decay rates of $B\to D^{(*)}\tau\nu$ and $B\to D^{(*)}\ell\nu$, where $\ell=e,\mu$. If it is accounted for by new physics, there is no reason in general that the relevant neutrinos are, respectively, $\nu_\tau$ and $\nu_\ell$. We study whether the $\tau$ related rate could be enhanced by significant contributions to $B\to D^{(*)}\tau\nu_\ell$ from a class of operators in the Standard Model Effective Field Theory (SMEFT). We find the upper bounds from forbidden or rare meson decays imply that the contributions from the lepton flavor violating processes account for no more than about $4\%$ of the required shift. Yet, no fine-tuned flavor alignment is required for the new physics.  Searching for the related high-$p_T$ process $pp\to\tau^\pm\mu^\mp$ can at present put a lower bound on the scale of the lepton flavor violating new physics that is  a factor of $2.2$ weaker than the bound from meson decays. An exception to our conclusion arises from a specific combination of scalar and tensor SMEFT operators.
%\end{abstract}

%\end{titlepage}
%%%%%%%%%%%%%%%%%%%%%%%%%
\newpage

%\centerline{\bf $b\to c\tau\bar\nu_{e,\mu}$ contributions to $R(D^{(*)})$}

%\vspace{6 pt}
%%%%%%%%%%%%%%%%%%
\section{Introduction}
Within the Standard Model (SM), the electroweak interactions of the leptons are flavor universal. Violation of lepton flavor universality arises from Yukawa interactions, that are negligible in this context, and from phase space effects, which are calculable.  A test of the SM prediction of lepton flavor universality is provided by the ratios
\beq
R(D^{(*)})\equiv\frac{\Gamma(B\to D^{(*)}\tau\bar\nu)}{\Gamma(B\to D^{(*)}\ell\bar\nu)},\ \ \ (\ell=e\ {\rm or}\ \mu).
\eeq
The SM predictions, derived by naive averaging \cite{HFLAV:2019otj} over the results reported in Refs. \cite{Bigi:2016mdz,Bernlochner:2017jka,Bigi:2017jbd,Jaiswal:2017rve}, are
\beqa
R(D)&=&0.299\pm0.003,\no\\
R(D^*)&=&0.258\pm0.005.
\eeqa
The current world averages for $R(D)$ and $R(D^*)$, combining the results reported in Refs. \cite{Abdesselam:2019dgh,Aaij:2017deq,Aaij:2017uff,Hirose:2017dxl,Hirose:2016wfn,Aaij:2015yra,Huschle:2015rga,Lees:2013uzd,Lees:2012xj} are as follows \cite{HFLAV:2019otj}:
\beqa
R(D)&=&0.340\pm0.030,\no\\
R(D^*)&=&0.295\pm0.014.
\eeqa
The difference of the experimental measurements from the SM predictions corresponds to about $3.1\sigma$ (p-value of $2.7\times10^{-3}$). We thus aim to explain
\beq
R(D^{(*)})/R(D^{(*)})^{\rm SM}\approx1.14\pm0.05.
\eeq

The quark transition via which the $B\to D^{(*)}\tau\nu$ proceeds is $b\to c\tau\nu$. Note, however, that the flavor of the neutrino is, of course, unobservable. It could be $\nu_\tau$, in which case the process respects the accidental lepton flavor symmetry of the SM. There is no reason, however, that the symmetry is respected by new physics, particularly when the new physics violates lepton flavor universality, so that the neutrino could also be $\nu_\mu$ or $\nu_e$ or some combination of the three flavors. This possibility has been discussed very little in the literature (for an exception, see \cite{Feruglio:2016gvd}), and we aim to fill in this gap. We ask three main questions:
\begin{itemize}
\item Could the $R(D^{(*)})$ puzzle be solved via new physics contributions to $b\to c\tau\nu_\ell$ with $\ell=e,\mu$?
\item If not, how precise should the alignment of $\nu$ with $\nu_\tau$ be?
\item What is the sensitivity of the high-$p_T$ experiments at the LHC to such lepton flavor violating new physics?
\end{itemize}

The plan of this paper is as follows. In Section \ref{sec:smeft}, we present the theoretical framework of the SM effective field theory (SMEFT), within which we carry out our analysis, and estimate the size of the dimension-six operators that can explain the $R(D^{*})$ puzzle. In Section \ref{sec:bounds}, we obtain bounds on these operators from various processes to which they contribute. In Section \ref{sec:collider} we explore the reach of current and future collider experiments to probe the dimension-six terms with searches of lepton flavor violating di-lepton final states. We summarize our conclusions in Section \ref{sec:conclusions}. A discussion of additional operators is given in Appendix \ref{app:other}.

%%%%%%%%%%%%%%%%%%
\section{\boldmath $R(D^{(*)})$ in the SMEFT}
\label{sec:smeft}
We assume that the new physics contributions originate at a scale $\Lambda\gg v$, and consider the following two terms in the SMEFT Lagrangian \cite{Feruglio:2016gvd}:
\beq\label{eq:smeft}
{\cal L}_{\rm NP}=\frac{C_1^{ilkm}}{\Lambda^2}(\overline{L_i}\gamma_\sigma L_l)(\overline{Q_k}\gamma^\sigma Q_m)
+\frac{C_3^{ilkm}}{\Lambda^2}(\overline{L_i}\gamma_\sigma\tau^a L_l)(\overline{Q_k}\gamma^\sigma\tau^a Q_m),
\eeq
where $L$ is the $SU(2)$-doublet lepton field, $Q$ is the $SU(2)$-doublet quark field, and $i,l,k,m$ are flavor indices. For the sake of definiteness, and to avoid the strongest constraints from flavor changing neutral current (FCNC) processes, we take $i=\tau$, $k=s$, and $m=b$, while $l$ runs over $e,\mu,\tau$. Three comments are in order concerning our choices for the flavor, Lorentz and CP structures:
\begin{itemize}
\item The weakest constraints apply when $k=m=b$, in which case no FCNC in the down sector are generated. However, the contribution to the $b\to c\tau\bar\nu_\ell$ decay rate gets an extra suppression by $|V_{cb}|^2$ compared to the $k=s$ case. This brings the relevant new physics scale close to the electroweak scale (roughly, $(0.14)^{-1/4}m_W$), a situation that does not lend itself to an SMEFT analysis and requires a model dependent analysis instead (see, {\it e.g.} Ref.~\cite{Aloni:2017eny} for relevant simplified models). Bounds from $\Upsilon$ decays \cite{Abada:2015zea,Hazard:2016fnc,Patra:2022bih} are relevant for this scenario.
\item   The contributions from individual scalar or tensor operators are disfavored by their different contributions to $R(D)$ and $R(D^*)$, by the $B_c$ lifetime and by their modification of the differential decay rates with respect to the SM \cite{Blanke:2017qan,Li:2016vvp,Alonso:2016oyd,Celis:2016azn}. An exception to this statement is provided by a specific combination of scalar and tensor operators. We present further details on this issue in Appendix \ref{app:other}.
\item Given that we focus on lepton flavor violating operators, there is no interference with the SM operators. Thus, the measurements that we discuss are sensitive only to absolute values of Wilson coefficients, and not to their phase structure. 
\end{itemize}

We denote $C_{1,3}^{\tau lsb}$ by $C_{1,3}^l$. The $C_{1,3}^l$-dependent terms can be rewritten as follows:
\beqa\label{eq:c3c1terms}
\Lambda^2{\cal L}_{\rm NP}&=&(C_1^l+C_3^l)V_{is}V_{jb}^*(\overline{u_{Li}}\gamma^\mu u_{Lj})(\overline{\nu_\tau}\gamma_\mu\nu_l)\no\\
&+&(C_1^l-C_3^l)V_{is}V_{jb}^*(\overline{u_{Li}}\gamma^\mu u_{Lj})(\overline{\tau_L}\gamma_\mu l_L)\no\\
&+&(C_1^l-C_3^l)(\overline{s_{L}}\gamma^\mu b_{L})(\overline{\nu_\tau}\gamma_\mu\nu_l)\no\\
&+&(C_1^l+C_3^l)(\overline{s_{L}}\gamma^\mu b_{L})(\overline{\tau_L}\gamma_\mu l_L)\no\\
&+&2C_3^lV_{is}(\overline{u_{Li}}\gamma^\mu b_L)(\overline{\tau_L}\gamma_\mu \nu_l)\no\\
&+&2C_3^lV_{jb}(\overline{u_{Lj}}\gamma^\mu s_{L})(\overline{\tau_L}\gamma_\mu \nu_l)+{\rm h.c.}.
\eeqa

Thus, the SMEFT Lagrangian terms that contribute to $b\to c\tau\nu$ are
\beq
{\cal L}=\left(\frac{4G_FV_{cb}\delta_{l\tau}}{\sqrt{2}}+\frac{2C_3^lV_{cs}}{\Lambda^2}\right)(\overline{c_L}\gamma^\mu b_L)(\overline{\tau_L}\gamma_\mu\nu_l).
\eeq
We obtain:
\beq
\frac{R(D^{(*)})}{R(D^{(*)})^{\rm SM}}=1+\frac{\sqrt{2}}{G_F}{\cal R}e\left(\frac{V_{cs}}{V_{cb}}\frac{C_3^\tau}{\Lambda^2}\right)
+\frac{\sum_{\ell=e,\mu}|C_3^\ell|^2}{2G_F^2\Lambda^4}\left|\frac{V_{cs}}{V_{cb}}\right|^2,
\eeq
where we assume that the contribution of the term quadratic in $C_3^\tau$ is negligible compared to the term linear in $C_3^\tau$.

Thus, to account for the $R(D^{(*)})$ puzzle by purely $b\to c\tau\nu_\ell$, $\ell=e,\mu$, we need
\beq\label{eq:rdc3mu}
\left(\frac{\sum_{\ell=e,\mu}|C_3^\ell|^2}{\Lambda^4}\right)^{1/2}=(0.24\pm0.04)\ {\rm TeV}^{-2}=\frac{1}{[(2.0\pm0.2)\ {\rm TeV}]^2}.
\eeq
On the other hand, to account for the $R(D^{(*)})$ puzzle by purely $b\to c\tau\nu_\tau$, we need
\beq
\frac{C_3^\tau}{\Lambda^2}=(0.046\pm0.016)\ {\rm TeV}^{-2}\approx\frac{1}{[(4.7\pm0.8)\ {\rm TeV}]^2}.
\eeq
%

%%%%%%%%%%%%%%%%%%%%%%%%
\section{Bounds on $C_3^\ell$}
\label{sec:bounds}
If the $R(D^{(*)})$ puzzle is accounted for by purely $b\to c\tau\bar\nu_\ell$, Eq. (\ref{eq:rdc3mu}) implies that we need $|C_3^\ell|/\Lambda^2\sim1/(2\ {\rm TeV})^2$. Eq. (\ref{eq:c3c1terms}) implies that the $C_3^\ell$ term contributes, via four fermi operators with the flavor structures $\bar s b\bar\tau\ell$ and $b\bar s\bar\nu_\tau\nu_\ell$, to various flavor changing neutral current and lepton flavor violating processes which are forbidden in the SM. In this section, we obtain the constraints from the experimental upper bounds on such processes.

\begin{itemize}
\item $B_s\to\tau^\pm\mu^\mp$.
\end{itemize}
The $B_s\to\tau^\pm\mu^\mp$ decay rate is given by
\beq
\Gamma(B_s\to\tau^+\mu^-)=\frac{|C_1^\mu+C_3^\mu|^2}{\Lambda^4}\frac{f_{B_s}^2 m_\tau^2 m_{B_s}}{64\pi}
\left(1-\frac{m_\tau^2}{m_{B_s}^2}\right)^2.
\eeq
The experimental upper bound \cite{Aaij:2019okb},
\beq
{\cal B}(B_s\to\tau^\pm\mu^\mp)<4.2\times10^{-5},
\eeq
implies
\beq\label{eq:bstaumubound}
\frac{|C_1^\mu+C_3^\mu|}{\Lambda^2}<0.073\ {\rm TeV}^{-2}.
\eeq

\begin{itemize}
\item $B^+\to K^+\tau^+\mu^-$.
\end{itemize}
The $B^+\to K^+\tau^+\mu^-$ branching ratio is given by
\beq
{\cal B}(B^+\to K^+\mu^-\tau^+)=8.2\times10^{-3}\ {\rm TeV}^4\times\frac{|C_1^\mu+C_3^\mu|^2}{\Lambda^4}.
\eeq
The experimental upper bound \cite{Lees:2012zz},
\beq
{\cal B}(B^+\to K^+\mu^-\tau^+)<2.8\times10^{-5},
\eeq
implies
\beq\label{eq:bfaclmutau}
\frac{|C_1^\mu+C_3^\mu|}{\Lambda^2}<0.058\ {\rm TeV}^{-2}.
\eeq

\begin{itemize}
\item $B^+\to K^+\tau^+e^-$.
\end{itemize}
The $B^+\to K^+\tau^+e^-$ branching ratio is given by
\beq
{\cal B}(B^+\to K^+e^-\tau^+)=8.2\times10^{-3}\ {\rm TeV}^4\times\frac{|C_1^e+C_3^e|^2}{\Lambda^4}.
\eeq
The experimental upper bound \cite{Lees:2012zz},
\beq
{\cal B}(B^+\to K^+ e^-\tau^+)<1.5\times10^{-5},
\eeq
implies (see also Ref.~\cite{Bause:2021cna})
\beq\label{eq:bfacletau}
\frac{|C_1^e+C_3^e|}{\Lambda^2}<0.044\ {\rm TeV}^{-2}.
\eeq

\begin{itemize}
\item $B^+\to K^+\bar\nu_\tau\nu_\ell$.
\end{itemize}
The $B^+\to K^+\nu\bar\nu$ branching ratio, normalized to the SM rate, is given by
\beq
{\cal R}_{K\nu\bar\nu}\equiv\frac{{\cal B}(B^+\to K^+\nu\bar\nu)}{{\cal B}^{\rm SM}(B^+\to K^+\nu\bar\nu)}
=1+3.5\times10^{3}\ {\rm TeV}^4\times\frac{|C_1^\mu-C_3^\mu|^2+|C_1^e-C_3^e|^2}{\Lambda^4}.
\eeq
The experimental upper bound \cite{Lees:2013kla,Grygier:2017tzo},
\beq
{\cal B}(B^+\to K^+\nu\bar\nu)<1.6\times10^{-5},
\eeq
which corresponds to $R_{K\nu\bar\nu}\lsim4$, implies (see also Ref.~\cite{Bause:2020auq})
\beq
\frac{|C_1^\ell-C_3^\ell|}{\Lambda^2}<0.031\ {\rm TeV}^{-2}.
\eeq

From the upper bounds on $|C_1^\ell\pm C_3^\ell|$ we can obtain upper bounds on $C_3^\ell$ alone. Assuming that $C_1^\ell$ and $C_3^\ell$ are real, we obtain:
\beq
\frac{|C_3^\mu|}{\Lambda^2}<0.044\ {\rm TeV}^{-2},\ \ \ \frac{|C_3^e|}{\Lambda^2}<0.037\ {\rm TeV}^{-2}.
\eeq
Comparing to the requirement of Eq.~(\ref{eq:rdc3mu}), we conclude that the contributions from $b\to c\tau\nu_\ell$ ($\ell=\mu,e$) can account for, at most, 3.4\% of the deviation of the central value of $R(D^{(*)})$ from the SM value. % Maybe more accurately stated as O(3%) unless we do a more precise calculation?

%%%%%%%%%%
\subsection{Other processes}
We here list a few processes to which the operators of Eq.~(\ref{eq:c3c1terms}) contribute. Due to our specific choice of flavor structure in Eq.~(\ref{eq:smeft}), these contributions are suppressed and provide weak limits only. With a different flavor structure, however, they might provide relevant bounds.

\begin{itemize}
\item Forbidden top decays, such as $t\to q\tau\ell$ with $q=c,u$ and $\ell=\mu,e$. There are currently no upper bounds on these decay rates \cite{ParticleDataGroup:2020ssz}.
\item Forbidden tau decays, for two of which there are experimental upper bounds \cite{Belle:2007cio,BaBar:2006jhm}:
\beqa
{\cal B}(\tau\to\mu\pi^0)&<&1.1\times10^{-7},\no\\
{\cal B}(\tau\to e\pi^0)&<&8.0\times10^{-8}.
\eeqa
Compared to the allowed $\tau\to\nu_\tau \bar uq\ (q=d,s)$, with ${\cal B}(\tau\to \nu\bar uq)\sim0.65$, there is a CKM suppression of $|V_{ub}V_{us}|^2\sim7\times10^{-7}$. Thus, branching ratios of ${\cal O}(10^{-7})$ do not provide significant bounds.
\item Allowed $\tau$ decays: The ratio ${\cal B}(\tau\to K\nu)/{\cal B}(\tau\to\pi\nu)$ is modified because the contribution is only to the former. Both branching ratios are measured, with \cite{ParticleDataGroup:2020ssz}
\beq
R_{K/\pi}\equiv\Gamma(K^-\nu)/\Gamma(\pi^-\nu)=(6.44\pm0.09)\times10^{-2}.
\eeq
The SM gives \cite{BaBar:2009lyd}
\beq
R_{K/\pi}=\frac{f_K^2|V_{us}|^2}{f_\pi^2|V_{ud}|^2}\frac{(1-m_K^2/m_\tau^2)^2}{(1-m_\pi^2/m_\tau^2)^2}(1+\delta_{\rm LD}),
\eeq
where $\delta_{\rm LD}=(0.03\pm0.44)\%$ is the long distance correction, and $f_K/f_\pi=1.189\pm0.007$. The new contribution to $\tau\to K\nu$ is CKM suppressed by $|V_{ub}/V_{us}|^2\sim3\times10^{-4}$ compared to the SM contribution, and so it is well below the uncertainty from $\delta_{\rm LD}$ and cannot be constrained.
\item Allowed $D_s^+$ decays: The ratio ${\cal B}(D_s\to\tau\nu)/{\cal B}(D_s\to\mu\nu)$ is modified because the contribution is only to the former. Both branching ratios are measured, with \cite{Belle:2013isi}
\beq
R_{\tau/\mu}=\frac{{\cal B}(D_s\to\tau\nu)}{{\cal B}(D_s\to\mu\nu)}=10.73\pm0.88.
\eeq
The SM gives
\beq
R_{\tau/\mu}^{\rm SM}=\frac{m_\tau^2(1-m_\tau^2/m_{D_s}^2)^2}{m_\mu^2(1-m_\mu^2/m_{D_s}^2)^2}=9.76.
\eeq
Within our framework,
\beq
\frac{R_{\tau/\mu}}{R_{\tau/\mu}^{\rm SM}}=1+\frac{\sum_{\ell=e,\mu}|C_3^\ell|^2}{8G_F^2\Lambda^4}\left|\frac{V_{cb}}{V_{cs}}\right|^2.
\eeq
Given the suppression by $(1/8)|V_{cb}/V_{cs}|^2\sim2\times10^{-4}$, we learn that $R_{\tau/\mu}$ does not provide a significant bound.
\item Forbidden $D^0\to \tau^\pm e^\mp$ decays. There is currently no upper bound on these rates  \cite{ParticleDataGroup:2020ssz}. The new contributions are strongly suppressed by several factors, compared to the leading semileptonic decay: CKM suppression by either $|V_{cb}V_{us}/V_{cs}|^2\sim10^{-4}$ or $|V_{ub}|^2\sim10^{-5}$, phase space suppression by $(1-m_\tau^2/m_{D^0}^2)^2=0.0084$ and annihilation suppression by $f_D^2/m_D^2\sim0.01$. 
\item Forbidden $J/\psi$ decays \cite{BES:2004jiw,BESIII:2021slj}:
\beqa\label{eq:psiltau}
{\cal B}(J/\psi\to\mu^\pm\tau^\mp)&<&2.0\times10^{-6},\no\\
{\cal B}(J/\psi\to e^\pm\tau^\mp)&<&7.5\times10^{-8}.
\eeqa
Compared to the allowed $J/\psi\to \ell^+\ell^-$, with ${\cal B}(J/\psi\to \ell^+\ell^-)\sim0.06$, there is a CKM suppression $|V_{cb}V_{cs}|^2\sim1.6\times10^{-3}$. Furthermore \cite{Abada:2015zea,Hazard:2016fnc},
\beq
\frac{{\cal B}(J/\psi\to\tau^\pm\ell^\mp)}{{\cal B}(J/\psi\to \ell^+\ell^-)}\propto\left(\frac{m_{J/\psi}}{\Lambda}\right)^4.
\eeq
Thus, the bound on $\Lambda/\sqrt{|C_3^e|}$ is of ${\cal O}(6m_{J/\psi})$ and on $\Lambda/\sqrt{|C_3^\mu|}$ even weaker.
\end{itemize}

%%%%%%%%%%%%%%
\section{Collider searches}
\label{sec:collider}
The effective operators of Eq. (\ref{eq:smeft}) will contribute to the scattering process $pp\to\tau^\pm\mu^\mp X_h$, where $X_h$ stands for final hadrons. In this Section, we estimate the upper bound on $|C_1^\mu+C_3^\mu|/\Lambda^2$ that can be obtained at the LHC at present and in the future. (For related work, see \cite{Faroughy:2016osc,Greljo:2015mma,DiLuzio:2017chi,Buras:2014fpa,Greljo:2017vvb}
and, in particular, \cite{Choudhury:2019ucz,Angelescu:2020uug,Kumar:2020hpo}.)

We base our estimate on the ATLAS search for new physics in $pp\to\mu^+\mu^-$ (with up to one $b$-jet) at $\sqrt{s}=13$ TeV with $139\ {\rm fb}^{-1}$ of data \cite{ATLAS:2021mla}. (The phenomenological framework for the ATLAS analysis was suggested in Ref. \cite{Afik:2018nlr}.) Ref. \cite{ATLAS:2021mla} obtains a bound $\Lambda_{\mu\mu}>2.4\ {\rm TeV}$ on the scale that suppresses dimension-six $bs\mu\mu$ contact interaction. This limit is obtained in the analysis with a $b$-veto ($pp\to\mu^+\mu^-+0b$), by searching for events with high dimuon mass, $m_{\mu\mu}>1800\ {\rm GeV}$. A somewhat weaker bound, $\Lambda_{\mu\mu}>2.0\ {\rm TeV}$, is obtained in  the $b$-tag category ($pp\to\mu^+\mu^-+1b$) with $m_{\mu\mu}>1600\ {\rm GeV}$. We deduce from these bounds the reach of ATLAS for $\Lambda_{\tau\mu}$, the scale that suppresses dimension-six $bs\tau\mu$ contact interaction.

The bound on $\Lambda_{\mu\mu}$ is inferred from the upper bound on $\sigma_{\mu\mu}$, the $\mu\mu$ signal cross section. When comparing to it a search for $pp\to\tau^\pm_h\mu^\mp+0b$, one muon is replaced by a hadronically decaying tau-lepton. Given that ${\cal B}(\tau\to{\rm hadrons})\approx2/3$, then at similar energy and for $\Lambda_{\tau\mu}=\Lambda_{\mu\mu}$, the signal cross-sections fulfill 
\beq
\sigma_{\tau_h\mu}/\sigma_{\mu\mu}\approx2/3.
\eeq

The leading SM background at the high $m_{\mu\mu}$ is the Drell-Yan process to two muons, which is suppressed in a $\tau\mu$ final state selection. As concerns the Drell-Yan process to two tau-leptons, its contribution to the background is suppressed by demanding that one of the two tau-leptons decays hadronically (with branching ratio $\sim2/3$) and the other muonically (with branching ratio $\sim1/6$): 
\beq
\sigma_{Z/\gamma^*}^{\tau_h\mu}/\sigma_{Z/\gamma^*}^{\mu\mu}\approx2\times2/3\times1/6=2/9.
\eeq
The next most significant SM background is the top-quark contribution, composed of $t\bar t$, $Wt$ and $Wt\bar t$. In each of these, the two final leptons arise from two independent decay chains, {\it e.g.} $t\to l^+$ and $\bar t\to l^-$ in $t\bar t$ events. Thus,
\beq
\sigma_{t}^{\tau_h\mu}/\sigma_t^{\mu\mu}\approx2\times2/3=4/3.
\eeq

Defining $r_{Z/t}\equiv\sigma_{Z/\gamma^*}^{\mu\mu}/\sigma_t^{\mu\mu}$, we have 
\beq
\frac{\sigma^{\rm bkgd}_{\tau_h\mu}}{\sigma^{\rm bkgd}_{\mu\mu}}\approx\frac{(2/9)r_{Z/t}+4/3}{r_{Z/t}+1}.
\eeq
We thus estimate
\beq
\frac{(s/\sqrt{b})_{\tau_h\mu}}{(s/\sqrt{b})_{\mu\mu}}\approx\frac{2}{3}\sqrt{\frac{r_{Z/t}+1}{(2/9)r_{Z/t}+4/3}}.
\eeq
For $r_{Z/t}>1$, we have $(s/\sqrt{b})_{\tau_h\mu}/(s/\sqrt{b})_{\mu\mu}\gsim0.75$, which is our conservative estimate. Note that if $r_{Z/t}\gsim4$, the sensitivity to $\tau\mu$ is in fact stronger than to $\mu\mu$. 

Given our conservative estimate, and that the signal sensitivity is fixed by experiment, the observational significance in $\tau\mu$ would be the same as in $\mu\mu$ for $s_{\tau\mu}\sim(4/3)s_{\mu\mu}$. Since $s\propto1/\Lambda^4$, we expect that current data can put a lower bound of
\beq\label{eq:lhclmutau}
\Lambda_{\tau\mu}>2.2\ {\rm TeV}.
\eeq
If the HL-LHC achieves its target integrated luminosity, $L=4000\ {\rm fb}^{-1}$, and no signal is observed, the bound would be strengthened to
\beq\label{eq:hllhclmutau}
\Lambda_{\tau\mu}>3.3\ {\rm TeV},
\eeq
where we scaled according to $s/\sqrt{b}\propto\sqrt{L}$ since $s/\sqrt{b} \sim L \sigma_{s} / \sqrt{L \sigma_{b}}$.

The analogous bounds $\Lambda_{ee}>2.0(1.8)\ {\rm TeV}$ are also attained in Ref.~\cite{ATLAS:2021mla}, from the analysis of $pp\to e^+e^-+0b(1b)$ events with high dielectron mass, $m_{ee}>1900(1500)\ {\rm GeV}$. We deduce from these bounds the reach of ATLAS for $\Lambda_{\tau e}$, the scale that suppresses dimension-six $bs\tau e$ contact interaction. The analysis goes along the same lines as the $\Lambda_{\tau\mu}$ analysis above. Given the lepton flavor universality of the relevant SM interactions, there are only negligible differences in the background cross sections. Our conservative estimates imply that current data can put a lower bound of
\beq\label{eq:lhcletau}
\Lambda_{\tau e}>1.9\ {\rm TeV}.
\eeq
For the HL-LHC with $L=4000\ {\rm fb}^{-1}$, and no signal observed, the bound would be strengthened to
\beq\label{eq:hllhcletau}
\Lambda_{\tau e}>2.8\ {\rm TeV}.
\eeq

For the $\tau\ell+0b$ ($\ell=\mu,e$) final states, our framework predicts additional contributions to those coming from the $b\bar s\tau\ell$ contact interactions. In particular, there are also contributions from $u\bar c\tau\ell$ and $u\bar u\tau\ell$ contact interactions. These contributions are, however, CKM suppressed, the former by $|V_{us}V_{cb}|^2\sim10^{-4}$, and the latter by $|V_{us}V_{ub}|^2\sim10^{-6}$. Involving the valence $u$-quark, instead of the sea $s$ or $b$ quark, gains less than two orders of magnitude from the parton distribution function, so that these contributions can be safely neglected in our analysis.

The operators of Eq.~(\ref{eq:c3c1terms}) lead also to mono-$\tau$ signatures at the LHC. The ATLAS \cite{ATLAS:2018ihk,ATLAS:2021bjk} and CMS \cite{CMS:2018fza} experiments have searched for $\tau\nu$ resonances, and their results have been recasted to apply to the SMEFT operators of interest to us \cite{Greljo:2018tzh,Marzocca:2020ueu,Endo:2021lhi,Jaffredo:2021ymt}, yielding \cite{Marzocca:2020ueu} 
\beq\label{eq:lhcnutau}
\Lambda_{\tau\nu}>1.5\ {\rm TeV}.
\eeq
At the HL-LHC, with integrated luminosity of $L=3000\ {\rm fb}^{-1}$, the sensitivity would be strengthened to 2.3 TeV \cite{Endo:2021lhi} .

%%%%%%%%%%%%%%
\section{Conclusions}
\label{sec:conclusions}
While the $R(D^{(*)})$ puzzle concerns lepton universality violation, it concerns also lepton flavor violation. Our study aimed to answer several questions in this context: Can the $R(D^{(*)})$ puzzle be explained by purely lepton flavor violating new physics? Does the puzzle imply that the relevant new physics require non-generic flavor structure? Can searches for lepton flavor violation at the ATLAS and CMS experiments shed light on these questions?

We reached the following conclusions:  
\begin{itemize}
\item
Given that, to account for the central value of $R(D^{(*)})$, it is required that $|C_3^\ell|/\Lambda^2\simeq0.24\ {\rm TeV}^{-2}$, but other constraints require that $|C_3^\mu|/\Lambda^2<0.044\ {\rm TeV}^{-2}$, the contribution of $b\to c\tau\nu_\ell$, with $\ell=e,\mu$, to $R(D^{(*)})/R(D^{(*)})^{\rm SM}-1$ cannot exceed about 4\% of the required shift.
\item 
Given that, to account  for the central value of $R(D^{(*)})$, it is required that $|C_3^\tau|/\Lambda^2\simeq0.046\ {\rm TeV}^{-2}$, but phenomenological constraints require that $|C_3^\mu|/\Lambda^2<0.044\ {\rm TeV}^{-2}$, and $|C_3^e|/\Lambda^2\simeq0.037\ {\rm TeV}^{-2}$, we learn that no special alignment with the $\tau$-direction is needed to explain the $R(D^{(*)})$ puzzle.
\item
Conversely, if operators of the form 
\beq
\frac{C_3^l}{\Lambda^2}(\overline{L_\tau}\gamma_\sigma\tau^a L_l)(\overline{Q_s}\gamma^\sigma\tau^a Q_b)
\eeq
have $C_3^\tau$, $C_3^\mu$ and $C_3^e$ all of the same order of magnitude, $C_3^l/\Lambda^2\sim0.04\ {\rm TeV}^{-2}$, then the shift in $R(D^{(*)})$ will be dominated by a factor of order 30 by $C_3^\tau$, and all phenomenological constraints satisfied.
\item
Comparing Eqs. (\ref{eq:lhclmutau}) and (\ref{eq:hllhclmutau}) to Eq. (\ref{eq:bfaclmutau}), we conclude that future searches at the (HL-)LHC will have to achieve an improvement in sensitivity by a factor $\sim(2.5)13$ in order to compete with existing constraints from the $B$-factories on $|C_1^\mu+C_3^\mu|/\Lambda^2$. Comparing Eqs. (\ref{eq:lhcletau}) and (\ref{eq:hllhcletau}) to Eq. (\ref{eq:bfacletau}), we conclude that future searches at the (HL-)LHC will have to achieve an improvement in sensitivity by a factor $\sim(8.5)40$ in order to compete with existing constraints from the $B$-factories on $|C_1^e+C_3^e|/\Lambda^2$.
\end{itemize}

%%%%%%%%%%%%%%%%%%%%%%%%%%%%%%%%%%%%%%%
%%%%%%%%%%%%%%%%%%%%%%%%%%%%%%%%%%%%%%%
\appendix
\section{Additional operators}
\label{app:other}
In the context of $R(D^{(*)})$, the following three operators are considered, in addition to the operators of Eq.~(\ref{eq:smeft}):
\beqa\label{eq:osroslot}
O_{SR}^\ell&=&(\overline{Q_s}b_R)(\overline{\tau_R}L_\ell),\no\\
O_{SL}^\ell&=&(\overline{c_R}Q_b)(\overline{\tau_R}L_\ell),\no\\
O_T^\ell&=&(\overline{c_R}\sigma^{\mu\nu}Q_b)(\overline{\tau_R}\sigma_{\mu\nu}L_\ell).
\eeqa
At low energy, we can write the effective Hamiltonian terms that are relevant to $b\to c\tau\bar\nu_\ell$ transitions as follows (see {\it e.g.} \cite{Blanke:2018yud}):
\beq\label{eq:hatceff}
{\cal H}_{\rm eff}=2\sqrt{2}G_F V_{cb}[\hat C_S^\ell(\overline{c}b)(\overline{\tau_R}\nu_\ell)+\hat C_P^\ell(\overline{c}\gamma_5 b)(\overline{\tau_R}\nu_\ell)
+\hat C_T^\ell(\overline{c_R}\sigma^{\mu\nu}b_L)(\overline{\tau_R}\sigma_{\mu\nu}\nu_\ell)],
\eeq
where the $C^\ell_{S,P}$ coefficients are related to the $O_{SL}^\ell\pm O_{SR}^\ell$ operators. In other words, writing down the Wilson coefficients in the SMEFT as $C_X/\Lambda^2$, as in  Eq.~(\ref{eq:smeft}), the relation with the $\hat C_X$ coefficients of Eq.~(\ref{eq:hatceff}) is given by
\beq
\frac{C_X^\ell}{\Lambda^2}=1.32\ \hat C_X^\ell\ {\rm TeV}^{-2}=\frac{\hat C_X^\ell}{(0.87\ {\rm TeV})^2}.
\eeq

Unlike lepton flavor diagonal operators, the lepton flavor violating ones do not interfere with the SM contributions, independent of their Lorentz structure. Their contributions to various observables related to $b\to c\tau\bar\nu_\ell$ transitions are given by (see {\it e.g.} \cite{Blanke:2018yud})
\beqa
R(D)/R(D)^{\rm SM}&=& 1+1.09|\hat C_S^\ell|^2+0.75|\hat C_T^\ell|^2,\no\\
R(D^*)/R(D^*)^{\rm SM}&=& 1+0.05|\hat C_P^\ell|^2+16.3|\hat C_T^\ell|^2,\no\\
{\rm BR}(B_c\to\tau\nu)/{\rm BR}(B_c\to\tau\nu)^{\rm SM}&=& 1+18.5|\hat C_P^\ell|^2,
\eeqa
where the Wilson coefficients $C_X^\ell$ are given at the scale $m_b$.

Accounting for the $R(D^{(*)})$ puzzle by either of these operators is disfavored compared to the one we studied above:
\begin{itemize}
\item While $|\hat C_S^\ell|^2\approx0.13$ can account for $R(D)$, it leaves $R(D^*)=R(D^*)^{\rm SM}$.
\item While $|\hat C_P^\ell|^2\approx3$ can account for $R(D^*)$, it gives ${\rm BR}(B_c\to\tau\nu)\approx1$.
\item While $|\hat C_T^\ell|^2\approx0.009$ can account for $R(D^*)$, it leaves  $R(D)\simeq R(D)^{\rm SM}$.
\end{itemize}

A specific combination of operators may, however, account for $R(D)$ and $R(D^*)$ without violating the bound from $B_c\to\tau\nu$. The requirements are
\beq\label{eq:r2}
|\hat C_S^\ell(m_b)|\approx0.35,\ \ \ |\hat C_P^\ell(m_b)|\lsim1.5,\ \ \ |\hat C_T^\ell(m_b)|\approx0.09.
\eeq

This combination of parameters seems rather ad-hoc. A model that goes, however, in this direction is the $R_2$ leptoquark model of Refs.~\cite{Becirevic:2018afm,Feruglio:2018fxo} ($R_2(3,2)_{+7/6}$ is a scalar, color-triplet, $SU(2)$-doublet of hypercharge $+7/6$). It generates at the high scale $\Lambda$ of integrating out $R_2$ the following Wilson coefficients:
\beq
\hat C_S^\ell(\Lambda)=\hat C_P^\ell(\Lambda)=4\hat C_T^\ell(\Lambda).
\eeq
With $\Lambda\approx1$ TeV, the RGE modifies this relation into (see {\it e.g.} \cite{Blanke:2018yud})
\beq
\hat C_S^\ell(m_b)=\hat C_P^\ell(m_b)\approx8.1C_T^\ell(m_b),
\eeq
thus predicting
\beq\label{eq:rtwo}
r_{D/D^*}\equiv\frac{[R(D)/R(D)^{\rm SM}]-1}{[R(D^*)/R(D^*)^{\rm SM}]-1}\approx3.7.
\eeq
Two comments are in order:
\begin{itemize}
\item The original models, when fitting the data with a final $\nu_\tau$, find that $\hat C_S^\ell$ needs to be close to imaginary, to reduce the effect of interference terms. With a final $\nu_\ell$, the interference terms vanish identically, with no need for a special phase structure.
\item In the absence of interference terms, the predicted ratio (\ref{eq:rtwo}) is disfavored at the $3\sigma$ level by the current experimental range (we take into account the correlation between the measurements \cite{HFLAV:2019otj}):
\beq
r_{D/D^*}=0.96\pm0.92.
\eeq
\end{itemize}

Finally, let us mention that among the three operators of Eq.~(\ref{eq:osroslot}), only $O_{SR}^\ell$ generates FCNC processes in the down sector. In particular, the contribution of $|C_P^\mu|^2$ to the $\Gamma(B_s\to\tau^+\mu^-)$ decay is enhanced  by $[m_{B_s}^2/(m_\tau(m_b+m_s))]^2$ compared to the one of $|C_1^\mu+C_3^\mu|^2$. Thus the bound on $|C_P^\ell|/\Lambda^2$ is a factor $\sim m_{B_s}/m_\tau=3$ stronger than the bound of Eq.~(\ref{eq:bstaumubound}): $|C_P^\mu|/\Lambda^2<0.024\ {\rm TeV}^{-2}$ or, equivalently, $|\hat C_P^\mu|<0.018$, making its contribution to $R(D^*)$ negligible.

%%%%%%%%%%%%%%%%%
\subsection*{Acknowledgements}
%\noindent{\bf Acknowledgments:}
We thank Yoav Afik and Daniel Aloni for useful discussions.
SB is supported by the grants from the Israel Science Foundation (grant number 2871/19), the German Israeli Foundation (grant number I-1506-303.7/2019) and by the Yeda-Sela (YeS) Center for Basic Research.
YN is the Amos de-Shalit chair of theoretical physics, and is supported by grants from the Israel Science Foundation (grant number 1124/20), the United States-Israel Binational Science Foundation (BSF), Jerusalem, Israel (grant number 2018257), by the Minerva Foundation (with funding from the Federal Ministry for Education and Research), and by the Yeda-Sela (YeS) Center for Basic Research. 

%%%%%%%%%%%%%%%%%%%%%%%%

\end{document}